\documentclass[manuscript]{aastex}

\newcommand{\hessj}{\object{HESS~J0632+057}}

\shorttitle{Probing the nature of \hessj}
\shortauthors{Moritani et al.}

\begin{document}


\title{Probing the nature of the TeV $\gamma$-ray binary \hessj \ by monitoring Be disk variability}


\author{Y. MORITANI\altaffilmark{1,2}, 
A. T. OKAZAKI\altaffilmark{3},
A. C. CARCIOFI\altaffilmark{4},
A. IMADA\altaffilmark{5}, 
H. AKITAYA\altaffilmark{1}, \\
N. EBISUDA\altaffilmark{6},
R. ITOH\altaffilmark{6},
K. KAWAGUCHI\altaffilmark{6},
K. MORI\altaffilmark{6},
K. TAKAKI\altaffilmark{6},\\
I. UENO\altaffilmark{6} and
T. UI\altaffilmark{6}}


\altaffiltext{1}{Hiroshima Astrophysical Science Center, Hiroshima University, 1-3-1 Kagamiyama Higashi-Hiroshima, Hiroshima, Japan 739-8526}
\altaffiltext{2}{Research Fellow of Japan Society for the Promotion of Science}
\altaffiltext{3}{Faculty of Engineering, Hokkai-Gakuen University, Toyohira-ku, Sapporo, Japan 062-8605}
\altaffiltext{4}{Instituto de Astronomia, Geof\'{i}sica e Ci\^{e}ncias Atmosf\'{e}ricas, Universidade de S\~{a}o Paulo, Rua do Mat\~{a}o 1226, Cidade Universit\'{a}ria, S\~{a}o Paulo, SP 05508-090, Brazil}
\altaffiltext{5}{Okayama Astrophysical Observatory, National Astronomical Observatory of Japan, Asakuchi, Okayama 719-0232}
\altaffiltext{6}{Department of Physical Science, Hiroshima University, 1-3-1 Kagamiyama, Higashi-Hiroshima 739-8526, Japan}


\begin{abstract}
We report on monitoring observations of the TeV $\gamma$-ray binary \hessj , which were carried out to constrain the interaction between the Be circumstellar disk and the compact object of unknown nature, and provide for the first time high-dispersion (R$\gtrsim 50\,000$) optical spectra in the second half of the orbital cycle, from apastron through periastron.
The H$\alpha$, H$\beta$, and H$\gamma$ line profiles are found to exhibit remarkable short-term variability for $\sim$1 month after the apastron (phase 0.6--0.7), whereas they show little variation near the periastron.
These emission lines show ``S-shaped'' variations with timescale of $\sim$150 days, which is about twice that reported previously.
In contrast to the Balmer lines, no profile variability is seen in any \ion{Fe}{2} emission line.
We estimate the radii of emitting regions of the H$\alpha$, H$\beta$, H$\gamma$, and \ion{Fe}{2} emission lines to be $\sim$30, 11, 7, and 2 stellar radii ($\mathrm{R_*}$), respectively. 
The amplitudes of the line profile variations in different lines indicate that the interaction with the compact object affects the Be disk down to, at least, the radius of 7 $\mathrm{R_*}$ after the apastron.
This fact, together with little profile variability near the periastron, rules out the tidal force as the major cause of disk variability.
Although this leaves the pulsar wind as the most likely candidate mechanism for disk variations, understanding the details of the interaction, particularly the mechanism for causing a large-scale disk disturbance after the apastron, remains an open question.
\end{abstract}




\section{Introduction}
TeV $\gamma$-ray binaries are a subclass of binaries with a compact object, established in the 2000's \citep[e.g.,][for a recent review]{Dubus2013}.
These systems have a spectral energy distribution with a peak beyond 1 MeV and are variable at multi-wavelengths up to TeV energies.
There are 5 known binaries of this kind, all of which have either an O-type main-sequence star or a Be star with a circumstellar disk as the optical counterpart.
The nature of the compact object is unknown in all systems but one (\object{PSR~B1259-63}).
For such systems two competing scenarios have been proposed, based on the different nature of the compact object, and hence the different mechanism of high energy emissions.
One scenario assumes that the collision between a relativistic pulsar wind and a stellar wind and/or circumstellar disk produces strong shocks, where the high energy emission arises \citep[e.g.,][]{Maraschi1981}.
This pulsar wind scenario has successfully been applied to \object{PSR~B1259-63}.
On the other hand, the other scenario assumes the presence of an accreting black hole (or neutron star).
In this microquasar or accretion/ejection scenario, a large amount of mass transferred from a companion star powers relativistic jets, where $\gamma$-ray emission originates \citep[e.g.,][]{Levinson1996}.

\hessj \ ($\mathrm{06^h32^m59^s.25 \; +05^\circ48'01''.2}$) is a recently established TeV $\gamma$-ray binary \citep{Aharonian2007} comprised of a B0Ve star and a compact object of unknown nature.
The orbit is wide \citep[$\mathrm{P_{orb}=}315^{+6}_{-4}$ days;][] {Aliu2014} and highly eccentric \citep[$\mathrm{e=}0.83$;][]{Casares2012}.
The system shows correlated variability in the X-ray and TeV energy bands \citep{Aliu2014}, with two peaks in one orbital cycle: the primary outburst prior to apastron (orbital phase $\phi$ = 0.3--0.4) and the secondary outburst after it ($\phi \sim$ 0.6--0.9), with an X-ray dip in between.
This is puzzling, given that these outbursts and the dip occur when the compact object is orbiting very far from the Be star.
Understanding the mechanism for these phenomena is one of the key issues of this system.

The optical counterpart of \hessj , \object{MWC 148}, is established to be a classical Be star, i.e., a massive star with a geometrically thin, circumstellar disk, where Balmer and other emission lines originate \citep{Rivinius2013}.
\citet{Aragona2010} observed the H$\alpha$ emission line after apastron ($\phi \sim$ 0.7--0.8).
Their profiles over continuous 35 days showed an ``S-shaped'' variability with a period of $\sim$ 60 days.
\cite{Casares2012} reported the orbital modulation in the H$\alpha$ line profile parameters, likely caused by the interaction between the Be disk and the compact object.
However, the lack of observations in $\phi$=0.6--0.8 makes it difficult to  constrain the modulation after apastron.

In order to constrain the interaction between the Be disk and the compact object, we have been monitoring \hessj \ using various methods such as high-dispersion spectroscopy, photometry, and polarimetry.
In this paper we report the initial result of the line profile variabilities over 160 days from 2013 October to 2014 April, covering the second half of the orbital period.
In Sect. 2, we summarize our observations.
The results are described in Sect. 3.
In Sect. 4 we discuss the Be disk region affected by the compact object and the nature of the interaction.

\section{Observations}
Optical high-dispersion spectroscopic observations of \hessj \ were carried out on 12 nights from 2013 October 31 (phase 0.555) to 2014 March 11 (phase 0.969) at the Okayama Astrophysical Observatory with a 188 cm telescope and HIDES with fiber-feed system \citep{Kambe2013}.
Here, we calculated the orbital phase, $\phi$, taking the orbital period of 315 days from \citet{Aliu2014} and the origin (JD 2454857.5)  from \citet{Casares2012}, and setting the periastron phase to be 0 (In \citealt{Casares2012}, it is set to 0.967).
Note that, although \citet{Casares2012} determined the ephemeris using the orbital period of 321 days, the orbital parameters remain approximately unchanged \citep{Aliu2014}.
The observed spectra covered 4200--7400 \AA \ wavelength range, with S/N of $\sim150$ around H$\alpha$.
The typical wavelength resolution $R$ is $\sim 50\,000$.
The data were reduced in the standard way, using the IRAF\footnote{http://iraf.noao.edu/} echelle package.

High-dispersion spectra were also taken on two nights (2014 February 05, phase 0.861, and April 10, phase 0.064) using the Canada France Hawaii Telescope/ESPaDOnS \citep{Manset(2003)} in spectropolarimetric mode.
The spectra covered a wavelength range of 3700--10\,500 \AA \ with a resolving power of  $\sim 68\,000$.
We obtained reduced data using the Libre-Esprit/Upena\footnote{http://www.cfht.hawaii.edu/Instruments/Upena/} pipeline, provided by the instrument team. 
We rectified the normalized intensity by re-determining the continuum level around each line, in order to compare with HIDES data.
In this paper, we focus on the spectroscopic variability.
The polarimetric data will be discussed in a forthcoming paper.

In addition to spectroscopic data, $V$-, $R_C$-, and $I_C$-band  photometric data were obtained from 2013 September to  2014 May using HOWPol \citep{Kawabata2008} and HONIR \citep{Sakimoto2012} attached to Kanata 1.5m Telescope at Higashi-Hiroshima Observatory in Hiroshima, Japan.
We observed \hessj \ on 73, 71 and 10 nights in $V$, $R_C$, and $I_C$ bands, respectively, using HOWPol.
We used HONIR to observe the source on 8 nights, in these three filters.
The logs of the photometric and spectroscopic observations are given in Table \ref{tbl:log}.

\section{Results}
\begin{figure}
\epsscale{.78}
\plotone{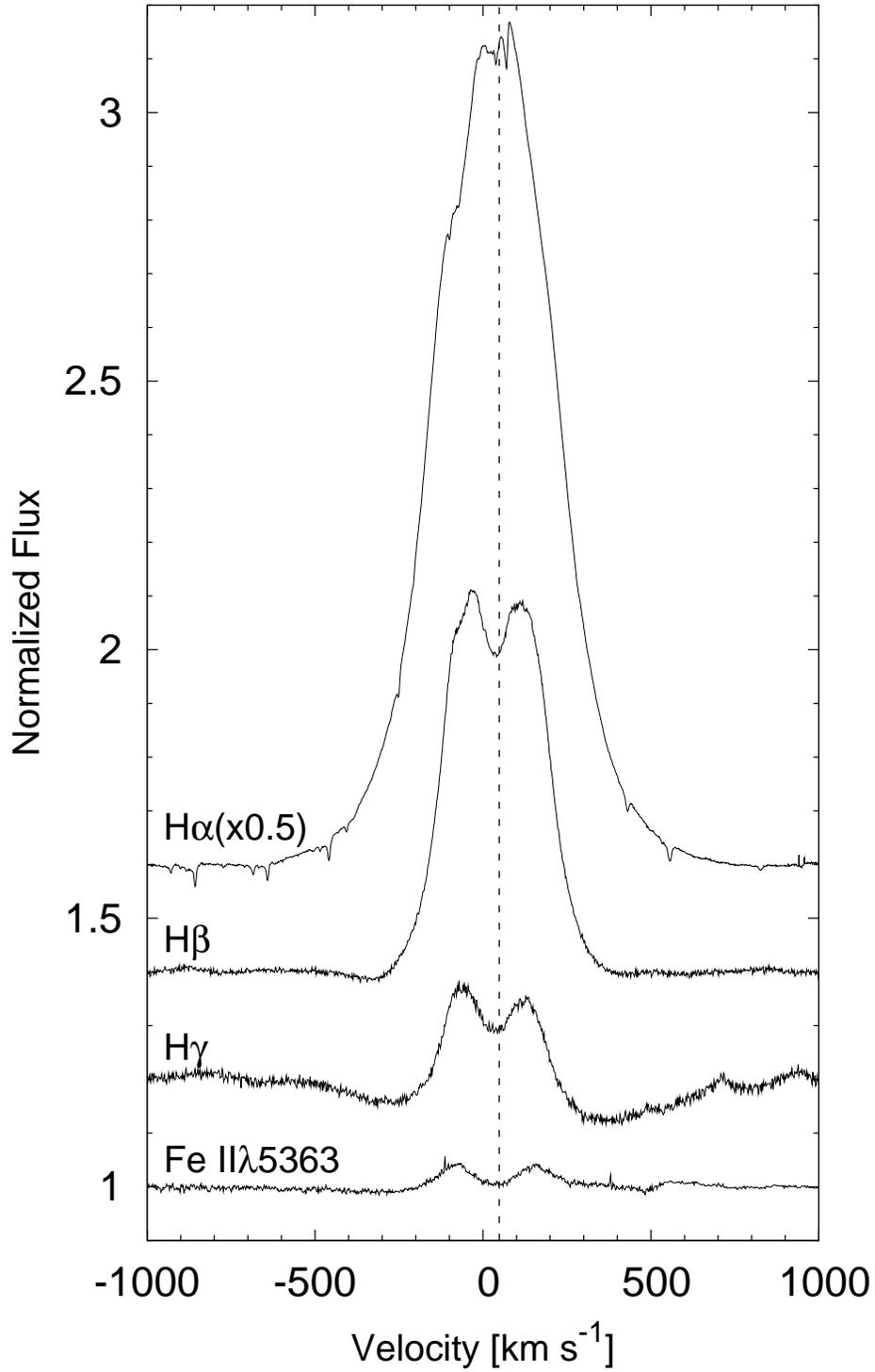}
\caption{
Averaged profiles of H$\alpha$, H$\beta$, H$\gamma$, and \ion{Fe}{2} $\lambda5363$ in the velocity reference frame.
For illustration purpose, there are vertical offsets between each profile, and the H$\alpha$ line flux is multiplied by 0.5.
The vertical dashed line indicates the systemic velocity \citep[$\mathrm{48.3\;km\;s^{-1}}$;][]{Casares2012}.
\label{fig:profave}
}
\end{figure}

Figure \ref{fig:profave} shows the averaged H$\alpha$, H$\beta$, and H$\gamma$ profiles.
In these profiles, the vertical scale of the H$\alpha$ line profile is half that of the other line profiles, because the flux is much stronger in the H$\alpha$.
The H$\alpha$ line is very strong ($EW \sim -30 \; \mathrm{\AA}$) and significantly asymmetric.
The profile exhibits a double peak at $\sim 0\;\mathrm{km\;s^{-1}}$ and $\sim70\;\mathrm{km\;s^{-1}}$, with the latter being brighter than the former, and a hump at $\mathrm{\sim-100}\;\mathrm{km\;s^{-1}}$ on the blue wing.
On the other hand, the averaged H$\beta$ line profile ($EW \sim -3.9 \; \mathrm{\AA}$) is rather symmetric, with a slightly stronger blue peak.
The H$\gamma$ line, which is on a broad absorption component, also exhibits a double-peaked profile with stronger blue peak.

\begin{figure}
\begin{center}
\begin{tabular}{cc}
H$\alpha$	& H$\beta$ \\
\includegraphics[scale=0.35]{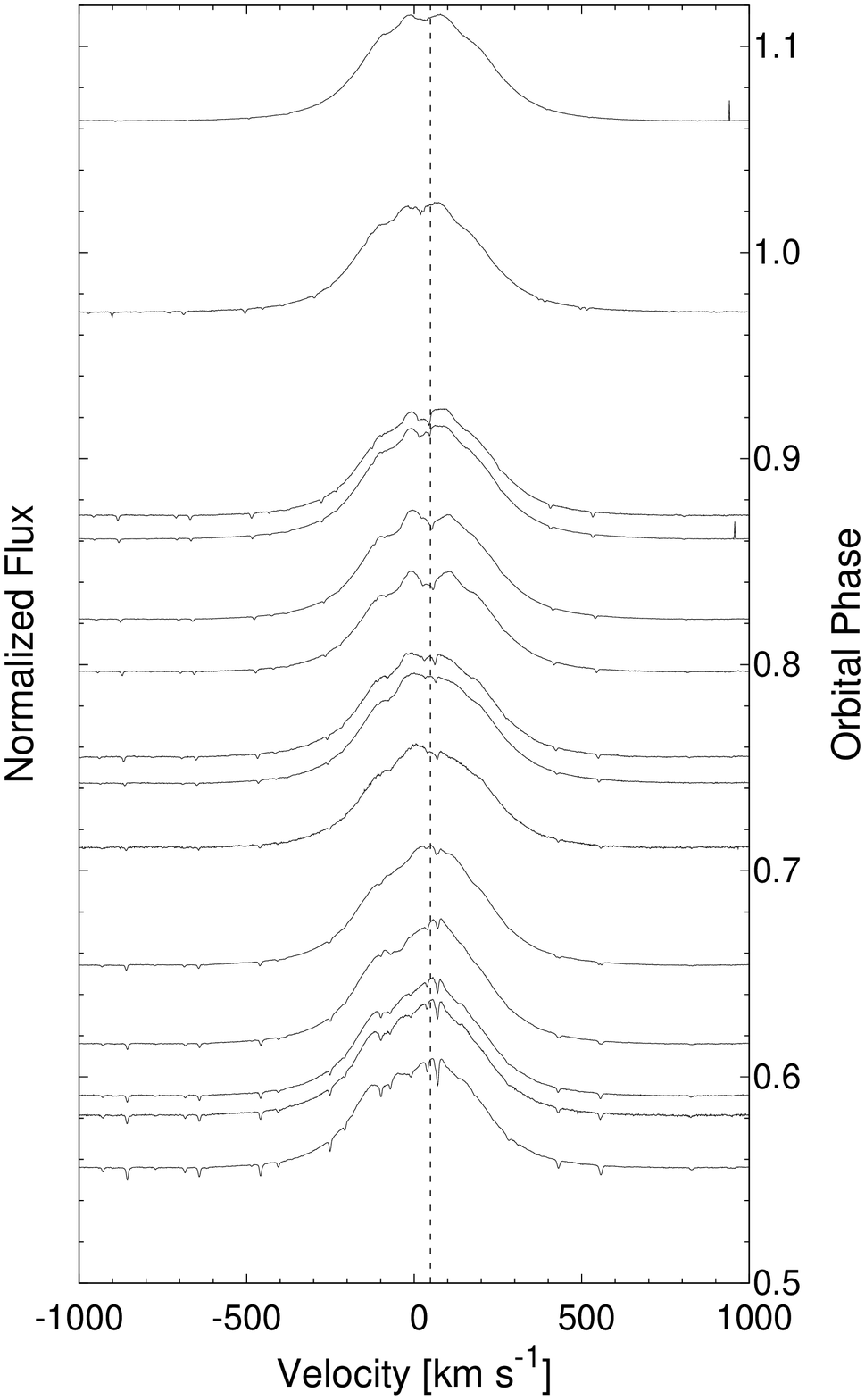}	& 
\includegraphics[scale=0.35]{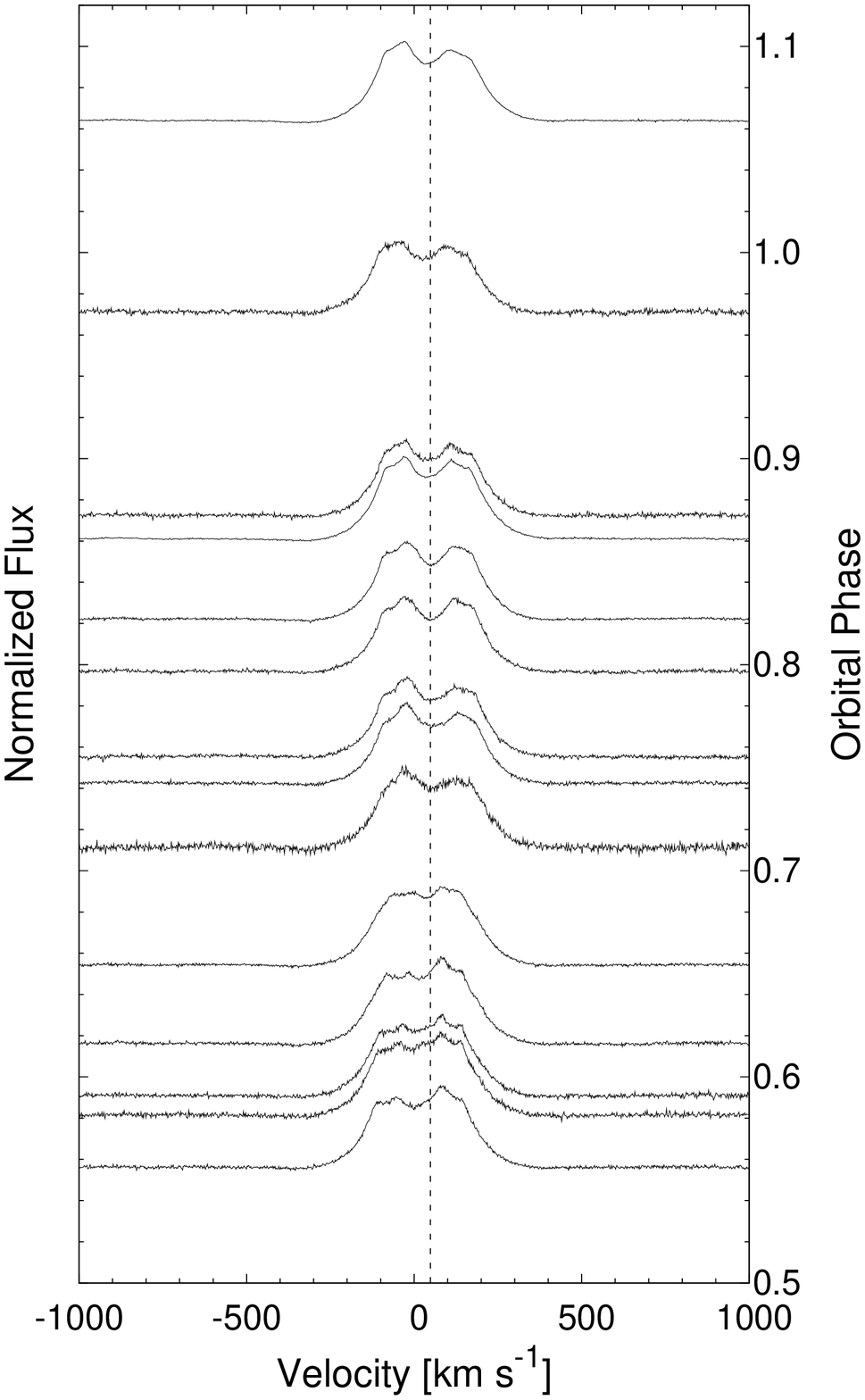}	\\
H$\gamma$	& \ion{Fe}{2} $\lambda$5363 \\
\includegraphics[scale=0.35]{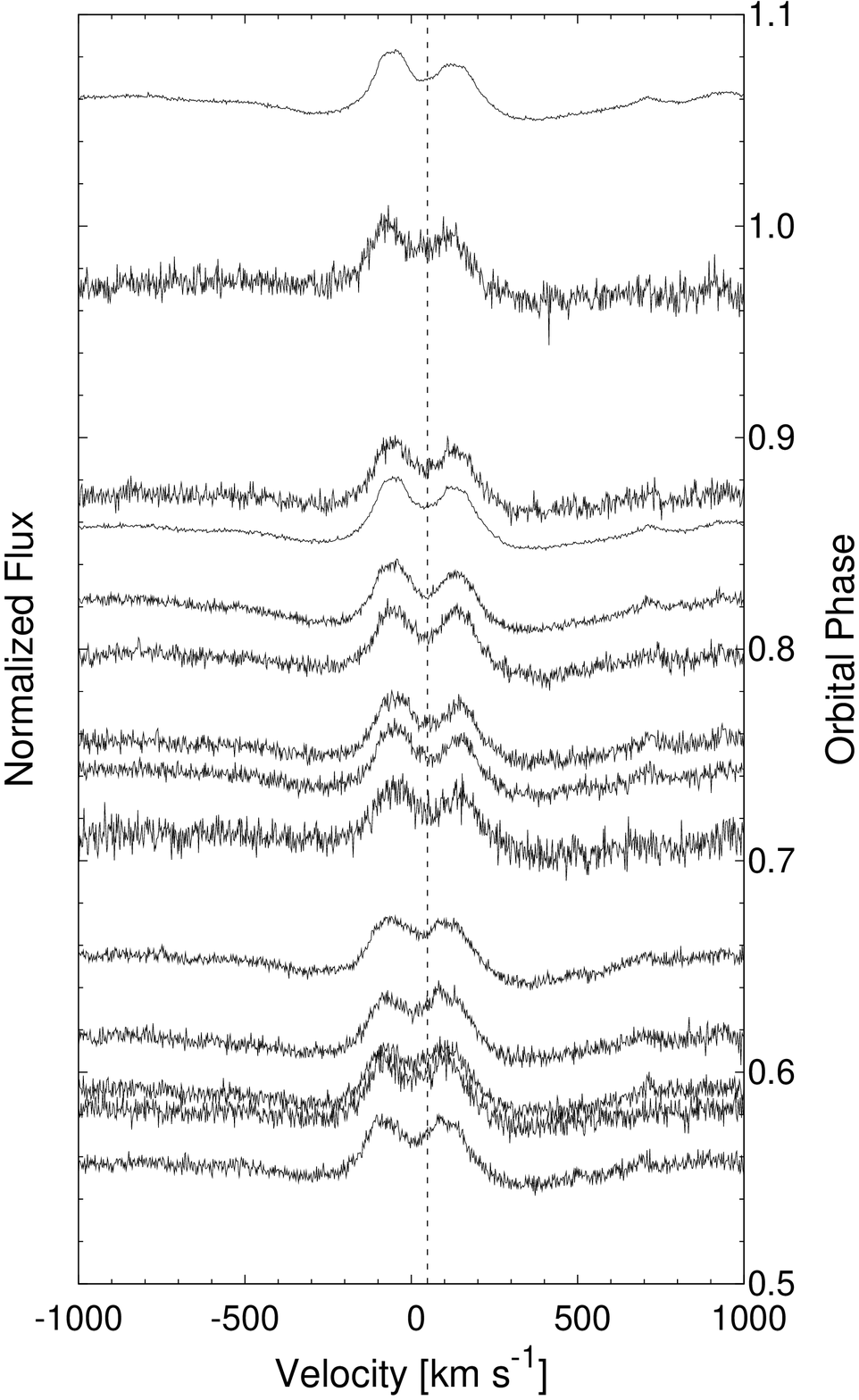}	& 
\includegraphics[scale=0.35]{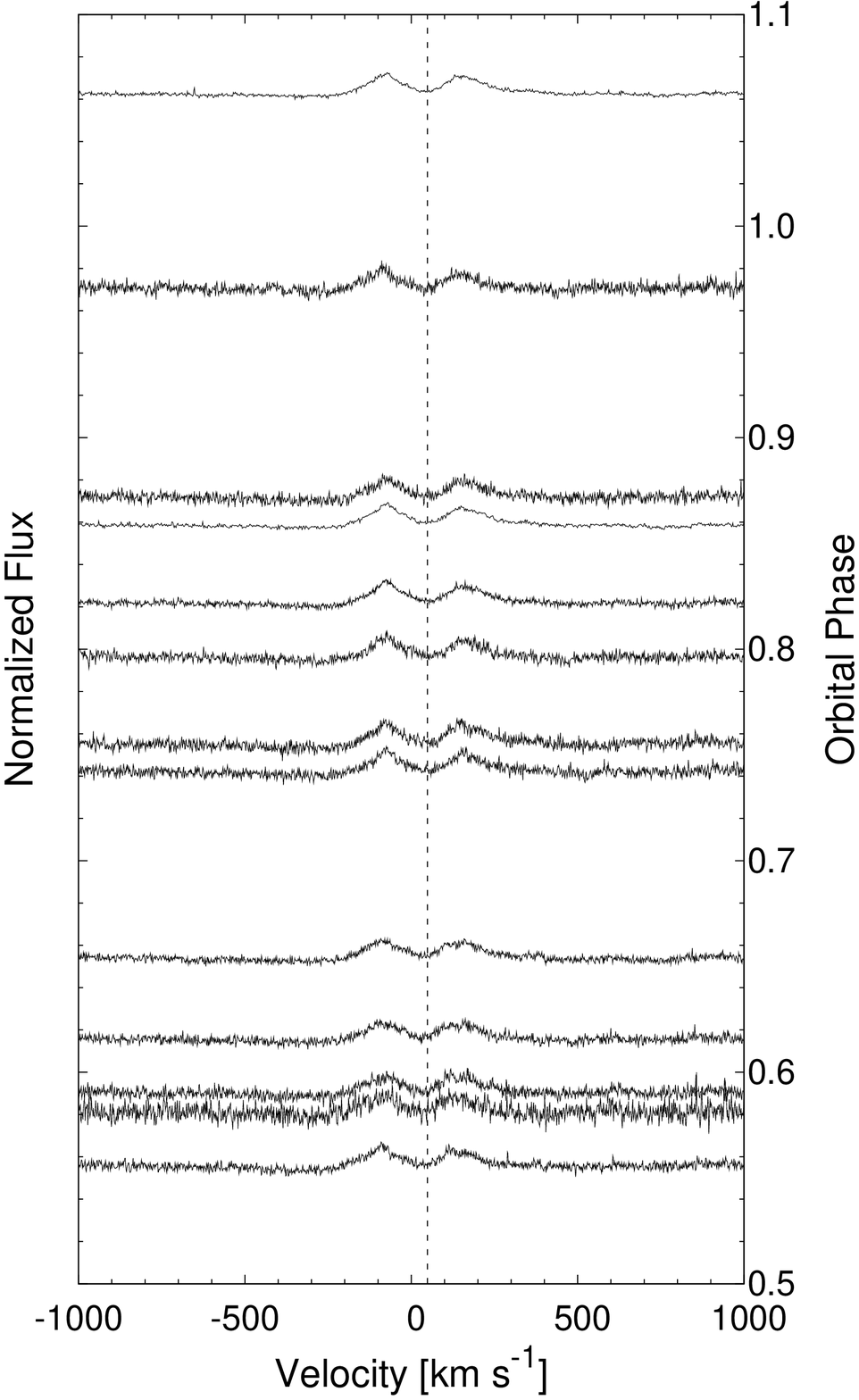}	\\
\end{tabular}
\caption{
Observed H$\alpha$, H$\beta$, H$\gamma$, and Fe\,II $\lambda5363$ line profiles.
Each profile is vertically shifted according to the orbital phase, labeled on the right axis.
\label{fig:prof_ind}
The vertical line is the same as Fig. \ref{fig:profave}.
}
\end{center}
\end{figure}

Balmer lines exhibited complicated line profile variabilities during the monitoring period.
Figure \ref{fig:prof_ind} displays all observed profiles of the H$\alpha$, H$\beta$, and H$\gamma$ lines, while Fig. \ref{fig:prof_diff} presents the time sequence of the residual spectra of these lines from the average.
The variation in the equivalent width, EW, the full width at half maximum, FWHM, and the centroid velocity, $V_c$, of the H$\alpha$ line profile are shown in Fig. \ref{fig:prof_param_ha}.
Note that FWHM and $V_c$ are measured by fitting the whole profile with a single Gaussian and the line wings with a Voigt profile, respectively.

In the long term ($\gtrsim 100$ days, from $\phi \sim 0.55-1$), the H$\alpha$ line changed from a red-enhanced profile to a rather symmetric profile, while the stronger peak of the H$\beta$ and H$\gamma$ lines changed from the red side to the blue side.
In the residual spectra of these lines, S-shaped variability is visible; the bright part at $\sim200\;\mathrm{km\;s^{-1}}$ migrates from the blue side to the red side at $\phi \sim 0.7$, and returns to the blue side at $\phi \sim 1.0$.
Simultaneously, the faint part at  $\sim -200\;\mathrm{km\;s^{-1}}$ migrates in the opposite direction.
At the same time, the maximum and minimum, respectively at $\sim100\;\mathrm{km\;s^{-1}}$ and $\sim-100\;\mathrm{km\;s^{-1}}$ in the residual spectra, migrate in a similar fashion. 
The period of these variations is estimated at $\sim 150$ days using Fourier analysis, which is about twice the $\sim 60$-day variation period reported by \citet{Aragona2010}.
We will discuss this in more detail in the next section.
On the other hand, in the short term ($\ll 100$ days), a bright hump appears at $\sim100\;\mathrm{km\;s^{-1}}$ in Balmer line profiles at $\phi \sim 0.6$.
This hump disappears by $\phi =0.65$ (see the third, forth, and fifth profiles from the bottom in Figs. \ref{fig:prof_ind} and \ref{fig:prof_diff}).
When the hump appears, EW(H$\alpha$) increases by $\sim 1 \mathrm{\AA}$  and FWHM(H$\alpha$) decreases by $\sim 10 \;\mathrm{km\;s^{-1}}$.
These long- and short-term variabilities respectively seem to occur in phase in different lines, although time lags between lines cannot be ruled out because of the low cadence of the observations.

Recently, \citet{Casares2012} reported the orbital modulation of the profile parameters of the H$\alpha$ line.
Their data cover the phase intervals of 0--0.6 and 0.8--1, using the orbital period of 321 days \citep{Bongiorno2011}.
If we take the same orbital period, our data spans the orbital phase 0.4--1, which overlaps with \citet{Casares2012} in the phase range 0.4--0.6 and 0.8--1 and provides for the first time the modulation data in the phase interval of 0.6--0.8.
In Fig. \ref{fig:prof_param_ha}, EW(H$\alpha$) seems to increase (phase 0.4--0.5) and then decrease back to the previous value in about 30 days (phase 0.5--0.6), as reported by \citet{Casares2012}.
Afterwards, it seems to fluctuate for about 60 days (phase 0.6--0.8), which could be associated with the variabilities mentioned above.
$V_c$(H$\alpha$), whose variation is thought to reflect the orbital motion of the Be star, shows a similar pattern to \citet{Casares2012} except just after the apastron and just before the periastron.
This difference in $V_c$ is possibly caused by the profile variabilities.
FWHM(H$\alpha$), on the other hand, shows a different variation pattern from \citet{Casares2012}.
It stays constant except around apastron (phase 0.45--0.55 for the period of 321 days) in this work, whereas in \citet{Casares2012} it showed a sinusoidal pattern.

Although many \ion{Fe}{2} lines are contaminated by neighboring lines, there are six less-contaminated lines ($\lambda=$5018, 5316, 5363, 5535, 6433, and 6456 \AA) that enable us to analyze the line profile variabilities.
In our analysis, we have used these six lines.
The averaged \ion{Fe}{2} $\lambda$5363 line profile in Fig. \ref{fig:profave} is shown as a representative \ion{Fe}{2} profile.
In contrast to the Balmer lines, the \ion{Fe}{2} emission lines show symmetric double-peaked profiles, as predicted for a rotating, axi-symmetric disk.
Figures \ref{fig:prof_ind} and \ref{fig:prof_diff} also display the time sequence of the observed and residual \ion{Fe}{2} $\lambda$5363 profiles, respectively. 
As seen in these figures, \ion{Fe}{2} lines exhibited no variation.
\citet{Arias2006} analyzed the spectra of several Be stars and confirmed that \ion{Fe}{2} emission lines arise from a disk region of radius of $2.0\pm0.8 \; \mathrm{R_*}$.
They also found empirical relationships between the projected rotational velocity of the central star ($V\sin i$) and the profile parameters.
Applying their relationships, we have derived $V\sin i$ to be $\sim 230-240\;\mathrm{km\;s^{-1}}$.

Finally, we estimate the radii of emitting regions of Balmer lines.
For the H$\alpha$ line, we use the mean peak separation of the last six profiles ($\phi \gtrsim 0.8$), because the averaged profile in Fig. \ref{fig:profave}, in which all our observed epochs were used, has too complicated features to determine the peak velocities.
For the H$\beta$ and H$\gamma$ lines, we use the averaged profiles.
As a result, we have the peak separations of the H$\alpha$, H$\beta$, and H$\gamma$ lines to be $90 \; \mathrm{km\;s^{-1}}$, $143\; \mathrm{km\;s^{-1}}$ and $176\; \mathrm{km\;s^{-1}}$, respectively.
Adopting $V\sin i$ derived above, we obtain the radii of the emitting region of these lines as $\sim 30$ $\mathrm{R_*}$ (H$\alpha$), 11 $\mathrm{R_*}$ (H$\beta$), and 7 $\mathrm{R_*}$ (H$\gamma$), where $R_*$ is the radius of the Be star.

Throughout our monitoring period (from apastron to periastron), optical brightness of \hessj \ stayed constant in the range of 9.05--9.32 ($V$) mag, 8.53--8.71 ($R_c$) mag, and 8.58--8.67 ($I_c$) mag, within the typical 1-$\sigma$ error of 0.05 mag.
The average values are $V=9.13 \pm 0.05$ mag, $R_C=8.58 \pm 0.06$ mag and $I_C=8.62\pm0.03$ mag.

\section{Discussion}
\subsection{The Disk Radius Affected by the Compact Object}
Although the monitoring was performed mostly after apastron, the line profiles exhibited remarkable variabilities.
Wide wavelength coverage revealed that the variations are seen not only in the H$\alpha$ line but also in the H$\beta$ and H$\gamma$ lines.
No variation in \ion{Fe}{2} lines indicates the inner part of the Be disk was kept undisturbed during the observing period.
This fact is in agreement with little variability in the optical brightness, which is thought to originate from the disk region within 2--3 $\mathrm{R_*}$ \citep{Rivinius2013}. 
The slight, but significant, variation in the H$\gamma$ line profile implies that the interaction with the compact object affects the Be disk down to, at least, a radius of 7 $\mathrm{R_*}$.

\subsection{S-shaped Variation}
The dynamical residual spectra of the Balmer lines showed an S-shaped variation (Fig. \ref{fig:prof_diff}), as \citet{Aragona2010} reported.
High-dispersion residual spectra clearly show that there are two pairs of peaks at $\sim -200/200$ and $\sim -100/100\;\mathrm{km\;s^{-1}}$.
The lower velocity peaks are the same as those of \citet{Aragona2010} (see their Fig. 3).
The period of the variation is $\sim$150 days, which is about twice the $\sim$ 60-day variation period \citep{Aragona2010}.

The S-shaped variation seen in many Be stars is thought to be caused by global disk oscillations.
They are mostly low-frequency $m=1$ oscillations, where $m$ is the azimuthal wave number \citep{Okazaki1991,Papaloizou1992}, but in eccentric binaries, it is also possible that $m>1$ oscillations are excited by the corresponding Fourier components of the tidal potential \citep[e.g.,][]{Artymowicz1994}.
If the period of observed variation is $\sim$ 150 days, about half the orbital period, it might be due to an $m=2$ oscillation mode in the Be disk, which could be excited by the $m=2$ Fourier component of the tidal potential of the compact object.

The difference of the periods between this work and \citet{Aragona2010} might be caused by the difference of the length and frequency of the observations.
Because we observed \hessj \ for about half an orbital cycle ($\sim 160$ days), variabilities of this timescale can be detected.
With low cadence of observation, typically once every 10--20 days, however, profile variabilities of shorter timescales are invisible.
On the contrary, the short monitoring epoch of \citet{Aragona2010} (35 days) would have hidden the variability with longer timescales ($>100$ days). 
Moreover, observations of this work and \citet{Aragona2010} have an interval of  $\sim$1800 days.
Therefore, there are possibilities that the oscillation mode has changed or that both oscillation modes exist.

\subsection{Short-Term Variability}\label{sec:Short}
In addition to the S-shaped variation, a remarkable variability was seen in the Balmer lines after apastron ($\phi$=0.60--0.65).
The short lifetime of this variability ($\lesssim 50$ days) indicates that it is temporarily caused by an external  force such as the tidal force and the ram pressure of the pulsar wind, because internal waves/oscillations have much longer lifetimes \citep[$\sim 1000$ days;][and references therein]{Okazaki1991}.
It is surprising, however, that such a remarkable variability appeared after apastron, when the interaction is expected to be weak because of the large distance between the Be disk and the compact object.
In order to understand this phenomenon, it is essential to constrain the starting phase and repeatability of the variability.

Comparison of the EW(H$\alpha$) between this work and \citet{Casares2012} implies the presence of a regular orbital modulation, but different absolute values due to different spectral resolutions make it difficult to quantitatively analyze it.
Given the orbital period close to one year, it is necessary to monitor over two or three more successive cycles, within the timescale of Be disk variability ($\lesssim 1000$ days), in order to cover the full orbital phase with the same spectral resolution.

\subsection{Implications for the Nature of the System}\label{sec:Imp}
Because the compact object is located far away from the Be star around apastron ($\sim 100 \; \mathrm{R_*}$), the gas in the Be disk at the radius of 7 $\mathrm{R_*}$ cannot be significantly affected by the tidal force of the compact object, whose strength is less than $10^{-4}$ of the gravity of the Be central star at this radius.
This leaves the interaction with the pulsar wind as the most likely mechanism for the Balmer line variability.

At the periastron passage, on the other hand, the compact object is thought to pass through the Be disk at $\sim 10 \; \mathrm{R_*}$.
Nonetheless, the H$\beta$ line, emitted from the disk region of the similar radius, showed no remarkable variation at $\phi=$0.06 (see the top profile in the upper right panel of Fig. \ref{fig:prof_ind}).
This fact suggests that either the tidal interaction is very weak in this system or the orbital period is longer than 315 days, so that the compact object had not passed the disk yet at the time of observation.
The former possibility could be realized if the Be disk rotates in the retrograde direction or is tilted by a large angle with respect to the orbital plane, where the disk gas and the compact object interact on a very short timescale.
Future observations at significantly later phases will distinguish between these two possibilities.

\subsection{Flip-Flop Scenario}
\begin{figure}
\begin{center}
\begin{tabular}{cc}
\vspace*{-10mm}
H$\alpha$	& H$\beta$ \\
\includegraphics[scale=0.5]{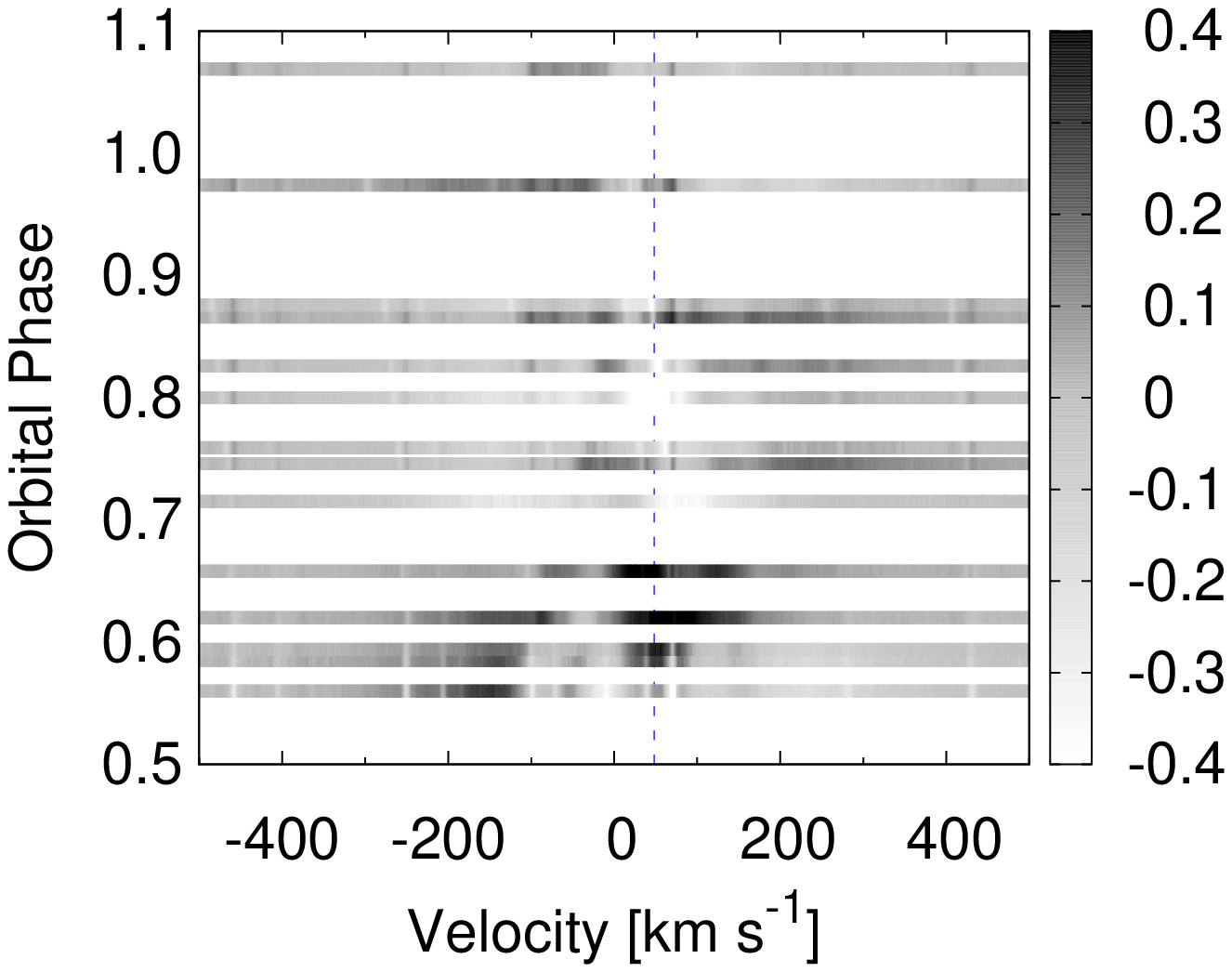}	& 
\includegraphics[scale=0.5]{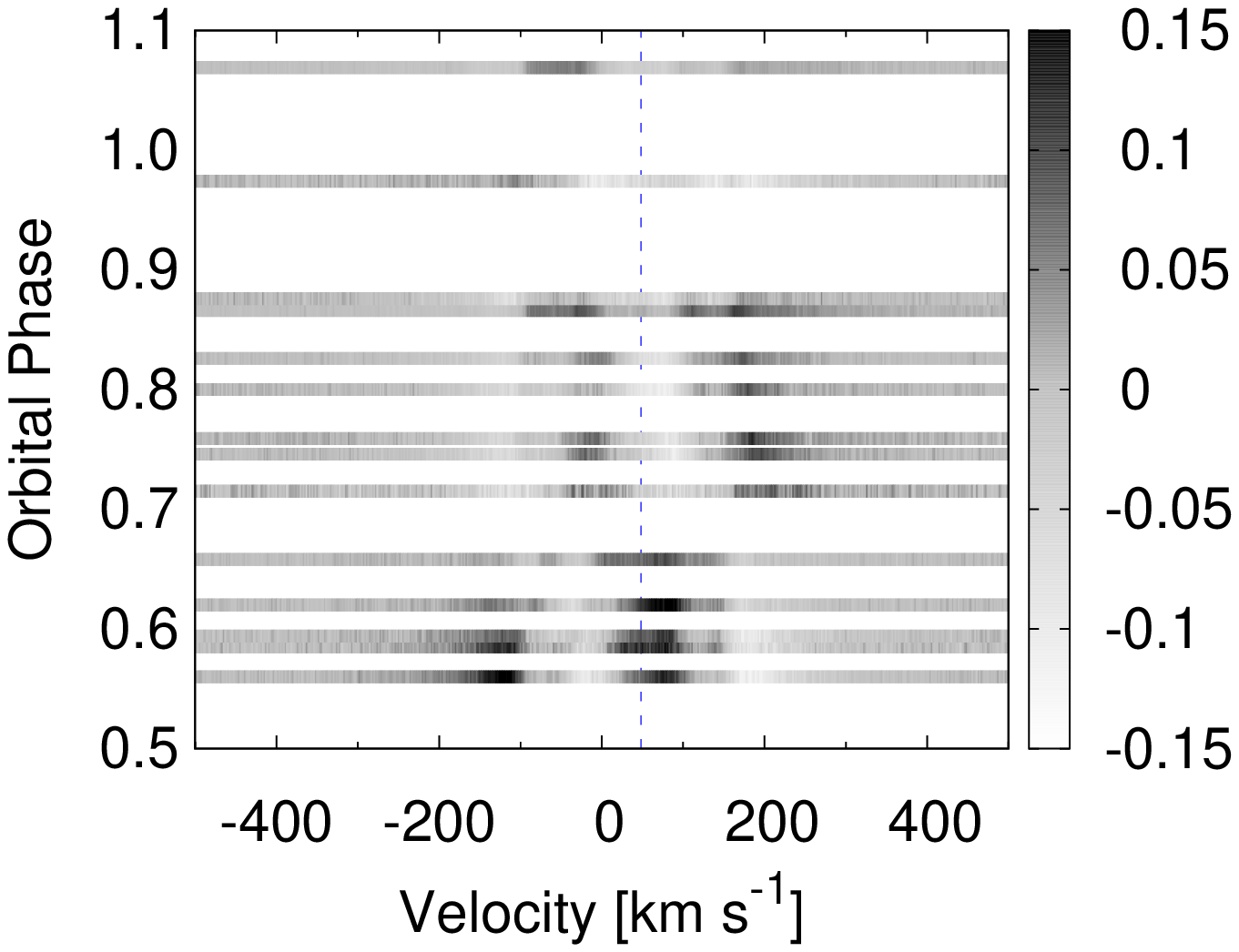}	\\
\vspace*{-10mm}
H$\gamma$	& Fe\,II $\lambda$5363 \\
\includegraphics[scale=0.5]{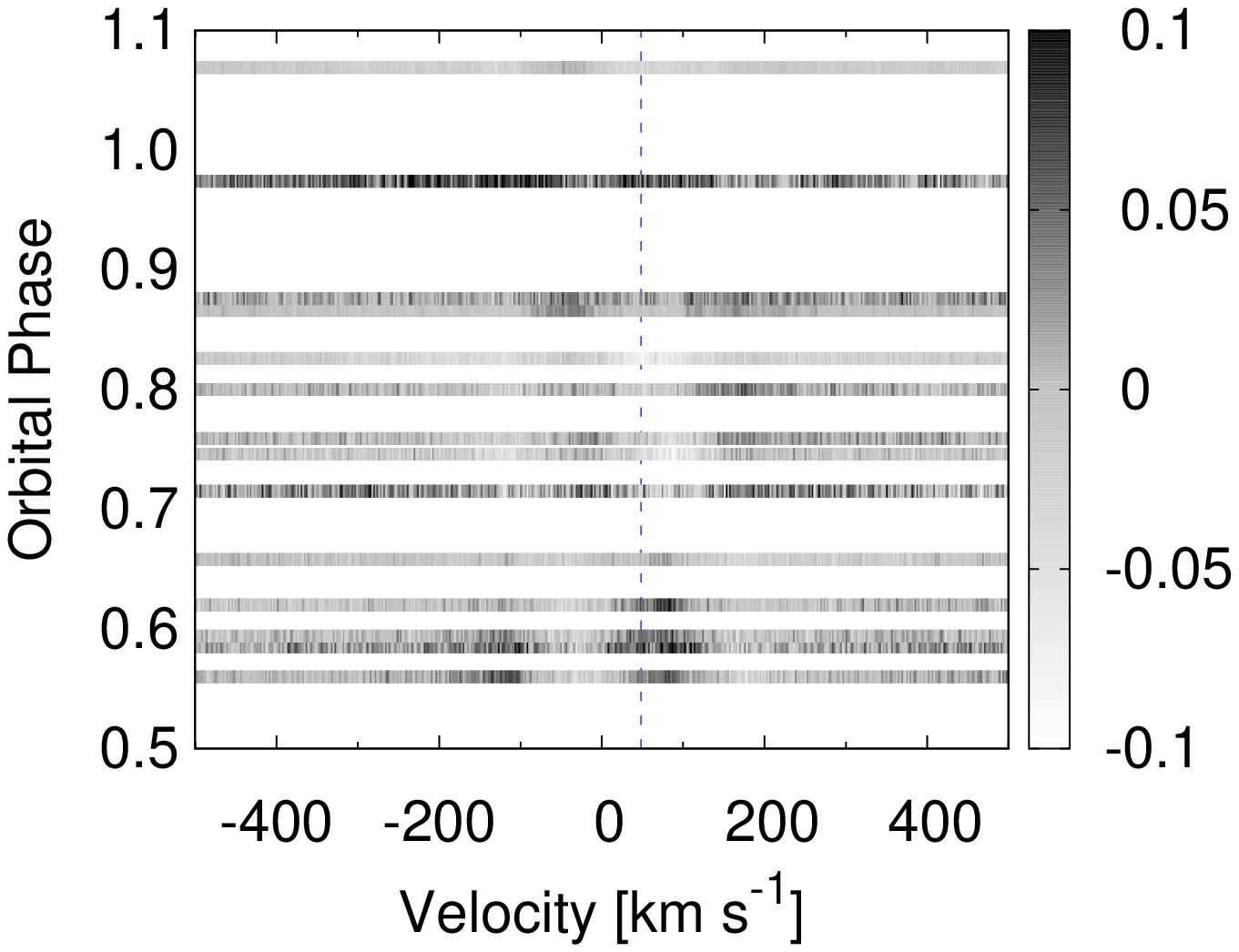}	& 
\includegraphics[scale=0.5]{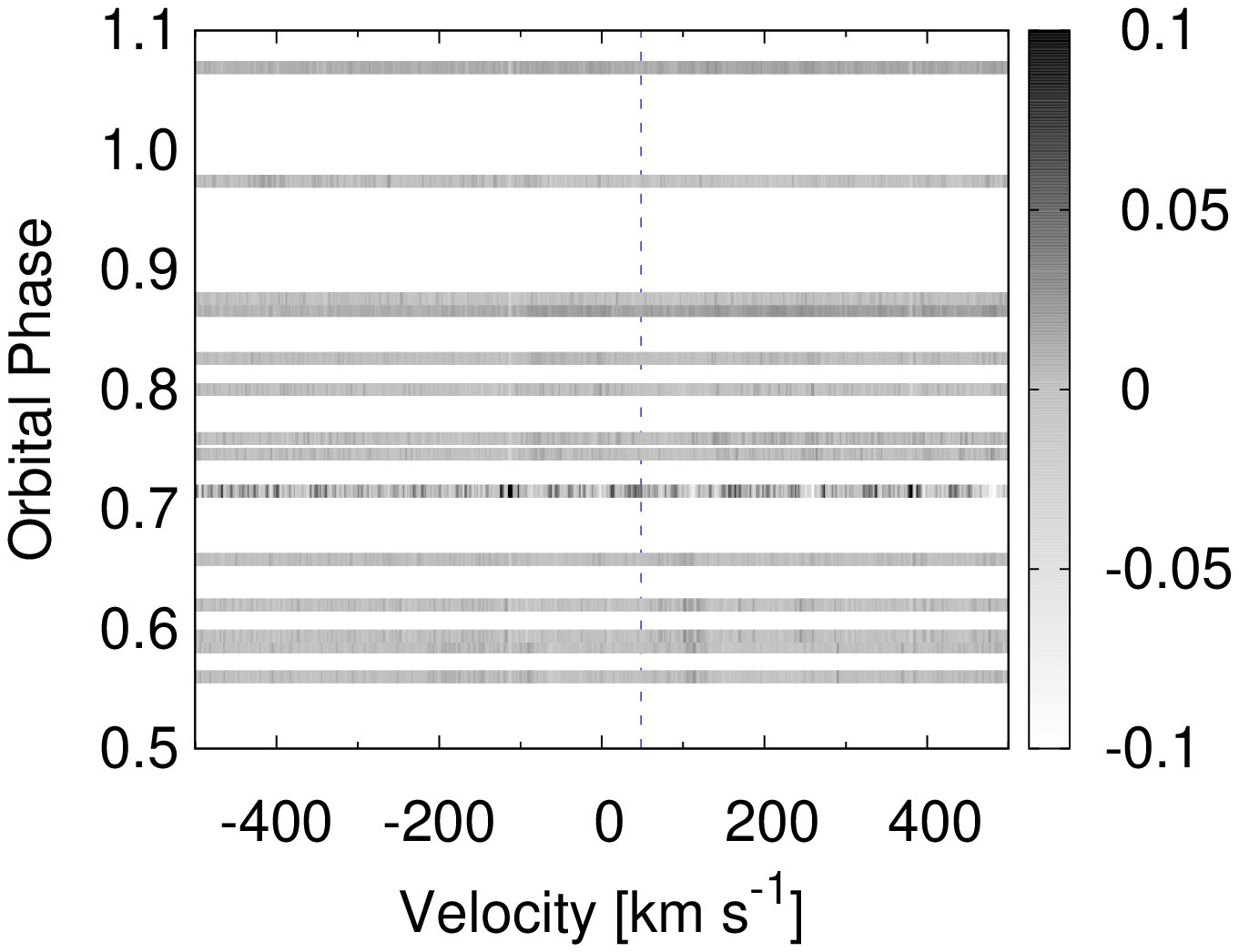}	\\
\end{tabular}
\caption{
Dynamical residual spectra of H$\alpha$, H$\beta$, H$\gamma$, and \ion{Fe}{2} $\lambda5363$ from the average profile.
The vertical line is the same as Fig. \ref{fig:profave}.
\label{fig:prof_diff}
}
\end{center}
\end{figure}

\begin{figure}
\epsscale{.80}
\plotone{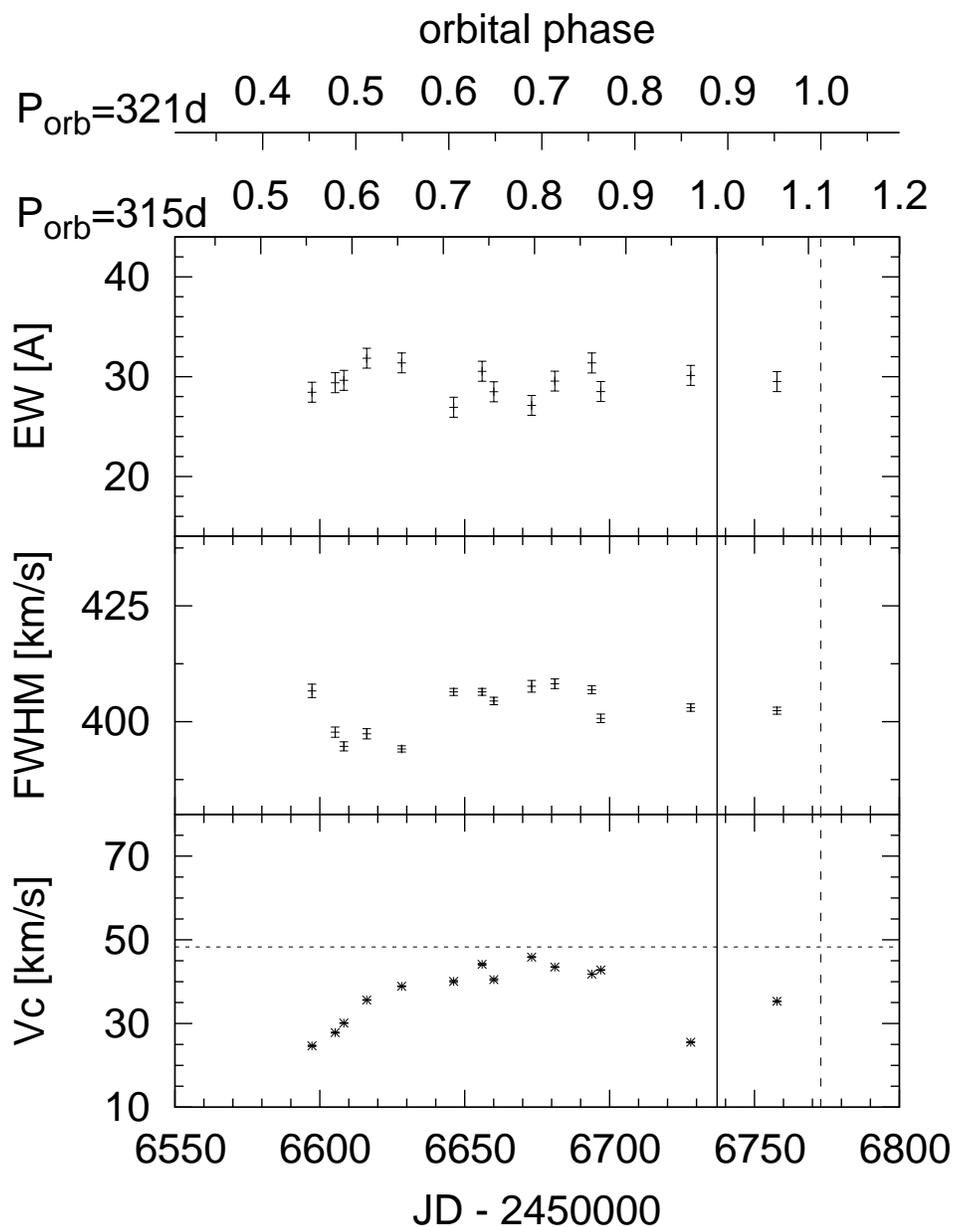}
\caption{
Variations of EW (top), FWHM (middle) and $V_c$ (bottom) of the H$\alpha$ line profile.
The orbital phase is annotated on the top of the figure, for two different orbital periods ($\mathrm{P_{orb}}$) of 315 and 321 days.
The vertical solid and dashed lines mark the periastron for the former and latter periods, respectively.
The horizontal line in the bottom panel indicates the systemic velocity \citep{Casares2012}.
\label{fig:prof_param_ha}
}
\end{figure}

As described in Sects. \ref{sec:Short} and \ref{sec:Imp}, our observations leave the pulsar wind model as the sole candidate for \hessj .
If the compact object is a pulsar, little line profile variation around periastron (the top two profiles in each panel of Fig. \ref{fig:prof_diff}) suggests that the pulsar wind is too weak to significantly affect the Be disk during these phases.
\citet{Torres2012} proposed a flip-flop scenario for another gamma-ray binary, \object{LS~I+61$^\circ$303}, where the pulsar is in a rotationally powered regime in the apastron, while it is in a propeller regime in the periastron \citep[see also][]{Papitto2012}.
In a flip-flop system, if the gas pressure of the Be disk overcomes the pulsar-wind ram pressure, the pulsar wind is quenched.
Because the Be disk of \hessj \ is estimated to be about three times larger than the binary separation at periastron, the compact object crosses a dense region of the disk near the periastron.
In such a situation, the strong gas pressure is likely to quench the pulsar wind and hence suppress high-energy emissions.

In the framework of the pulsar wind model, there are a few mechanisms that might explain the short-term episodic variability discussed above.
The Be disk is likely as large as the binary orbit because of no tidal truncation in highly eccentric, large orbit \citep{Okazaki2001}.
Because the disk density rapidly decreases with radius \citep[e.g.,][]{Carciofi2006}, the wind from the pulsar close to the outer part of the disk effectively changes its structure, giving rise to remarkable variability in emission lines arising from the disk outer part.
If the Be disk is misaligned with the orbital plane and a node happens to be in the direction corresponding to the phase of the variation ($\phi \sim 0.6-0.65$), the pulsar-wind effect on the Be disk will be the strongest in this phase interval.
After $\phi \sim 0.8$ the pulsar comes close to the denser disk region, which is not easily affected by the pulsar wind.
This might be a cause of the short-term episodic variation after aspastron.
It is not clear, however, how the pulsar wind can affect the inner disk at the radius of 7 $\mathrm{R_*}$.
Alternatively, the short-term post-apastron variations might be explained as the emission from the gas captured by the pulsar, if the pulsar wind is not strong enough to expel the surrounding gas.
This picture seems to fit well with the flip-flop model.
At any rate, observational investigation in the first half of the orbital cycle is needed in order to further test the pulsar-wind scenario.

The next periastron passage of \hessj \ will take place in 2015 December, according to the ephemeris of \citet{Aliu2014} (2016 January if the ephemeris is taken from \citealt{Casares2012}).
Observations covering this period will provide more clues to the complex interaction and the nature of the compact object in this puzzling $\gamma$-ray binary.



\acknowledgments

This paper is based on the observations taken at the Okayama Astrophysical Observatory.
We are grateful for Dr. Eiji Kambe to kindly observe \hessj .
This work was supported by Research Fellowships for the Promotion of Science for Young Scientists (YM, KT).
ATO acknowledges support by the JSPS Grant-in-Aid for Scientific Research (24540235) and a research grant from Hokkai-Gakuen Educational Foundation.
ACC acknowledges support from CNPq (grant 307076/2012-1). 



{\it Facilities:} \facility{OAO (HIDES-f)}, \facility{CFHT (ESPaDOnS)}, \facility{Kanata (HOWPol, HONIR)}.

\clearpage

\begin{table*}
\caption{Observation log.
\label{tbl:log}}
\begin{center}
\begin{tabular}{lccc}
\hline
\hline
\multicolumn{4}{c}{spectroscopic data} \\
\hline
\hline
& Date	& JD	&	phase	\\ \hline
OAO/HIDES	\\
& 2013.10.31	& 2456597.278	& 0.555		\\
& 2013.11.08	& 2456605.275	& 0.580		\\
& 2013.11.11	& 2456608.274	& 0.589		\\
& 2013.11.19	& 2456616.199	& 0.615		\\
& 2013.12.01	& 2456628.219	& 0.653		\\
& 2013.12.19	& 2456646.201	& 0.710		\\
& 2013.12.29	& 2456656.007	&0.741		\\
& 2014.01.02	& 2456660.047	& 0.754		\\
& 2014.01.15	& 2456673.095	& 0.795		\\
& 2014.01.23	& 2456681.094	& 0.821		\\
& 2014.02.08	& 2456696.968	& 0.871		\\
& 2014.03.11	& 2456728.008	& 0.969		\\
CFHT/ESPaDOnS \\
& 2014.02.05	& 2456693.843	& 0.861		\\
& 2014.04.10	& 2456757.772	& 0.064		\\
\hline
\hline
\multicolumn{4}{c}{photometric data} \\
\hline
\hline
& Filter	& JD	&	nights	\\ \hline
Kanata/HOWPol	\\
& $V$	& 2456551--2456795	& 73		\\
& $R_C$	& 2456551--2456795	& 71		\\
& $I_C$	& 2456722--2456795	& 10		\\
Kanata/HONIR\\
& $V$, $R_C$, $I_C$ 	& 2456743--2456769	& 8		\\
\hline
\end{tabular}    
\end{center}
\end{table*}

\clearpage

\clearpage

\clearpage


\clearpage


\end{document}